\documentclass[12pt]{article}
\usepackage{amssymb,amsfonts}
\usepackage{graphics,psboxit,amsmath}
\usepackage{subfigure}
\usepackage{graphicx}
\usepackage{verbatim}
\usepackage{color}
\usepackage{hyperref}
\hypersetup{colorlinks}

\definecolor{darkred}{rgb}{1,0,0}
\definecolor{darkgreen}{rgb}{0,0.5,0}
\definecolor{darkblue}{rgb}{0,0,1}
\definecolor{orange}{rgb}{1,0.5,0}
\definecolor{green}{rgb}{0,1,0}
\definecolor{purple}{rgb}{.5,0,1}

\hypersetup{ colorlinks,
linkcolor=darkred,
filecolor=darkgreen,
urlcolor=darkblue,
citecolor=darkblue,
linktocpage=true }

\definecolor{markcolor}{rgb}{.25,0,1}

\definecolor{markcolor2}{rgb}{1,0,0}

\definecolor{markcolor3}{rgb}{0,1,0}

\def\hybrid{\topmargin 0pt    \oddsidemargin 0.05in 
        \headheight 0pt \headsep 0pt
        \textwidth 16.3cm      
        \textheight 22,5cm       
        \marginparwidth .875in
        \parskip 5pt plus 1pt   \jot = 1.5ex}

\hybrid

\catcode`\@=11

\def\marginnote#1{}
\newcount\hour
\newcount\minute
\newtoks\amorpm
\hour=\time\divide\hour by60 \minute=\time{\multiply\hour by60
\global\advance\minute by-\hour}
\edef\standardtime{{\ifnum\hour<12 \global\amorpm={am}%
        \else\global\amorpm={pm}\advance\hour by-12 \fi
        \ifnum\hour=0 \hour=12 \fi
        \number\hour:\ifnum\minute<10 0\fi\number\minute\the\amorpm}}
\edef\militarytime{\number\hour:\ifnum\minute<10 0\fi\number\minute}

\def\draftlabel#1{{\@bsphack\if@filesw {\let\thepage\relax
   \xdef\@gtempa{\write\@auxout{\string
      \newlabel{#1}{{\@currentlabel}{\thepage}}}}}\@gtempa
   \if@nobreak \ifvmode\nobreak\fi\fi\fi\@esphack}
        \gdef\@eqnlabel{#1}}
\def\@eqnlabel{}
\def\@vacuum{}
\def\draftmarginnote#1{\marginpar{\raggedright\scriptsize\tt#1}}

\def\draft{\oddsidemargin -.5truein
        \def\@oddfoot{\sl preliminary draft \hfil
        \rm\thepage\hfil\sl\today\quad\militarytime}
        \let\@evenfoot\@oddfoot \overfullrule 3pt
        \let\label=\draftlabel
        \let\marginnote=\draftmarginnote
   \def\@eqnnum{(\theequation)\rlap{\kern\marginparsep\tt\@eqnlabel}%
\global\let\@eqnlabel\@vacuum}  }

\def\draft2{
        \def\@oddfoot{\sl preliminary draft \hfil
        \rm\thepage\hfil\sl\today\quad\militarytime}
        \let\@evenfoot\@oddfoot \overfullrule 3pt
        \let\label=\draftlabel
        \let\marginnote=\draftmarginnote
   \def\@eqnnum{(\theequation)\rlap{\kern\marginparsep\tt\@eqnlabel}%
\global\let\@eqnlabel\@vacuum}  }

\def\preprint{\twocolumn\sloppy\flushbottom\parindent 2em
        \leftmargini 2em\leftmarginv .5em\leftmarginvi .5em
        \oddsidemargin -.5in    \evensidemargin -.5in
        \columnsep .4in \footheight 0pt
        \textwidth 10.in        \topmargin  -.4in
        \headheight 12pt \topskip .4in
        \textheight 6.9in \footskip 0pt
        \def\@oddhead{\thepage\hfil\addtocounter{page}{1}\thepage}
        \let\@evenhead\@oddhead \def\@oddfoot{} \def\@evenfoot{} }

\def\numberbysection{\@addtoreset{equation}{section}
        \def\theequation{\thesection.\arabic{equation}}}

\def\underline#1{\relax\ifmmode\@@underline#1\else
        $\@@underline{\hbox{#1}}$\relax\fi}

\def\titlepage{\@restonecolfalse\if@twocolumn\@restonecoltrue\onecolumn
     \else \newpage \fi \thispagestyle{empty}\c@page\z@
        \def\thefootnote{\fnsymbol{footnote}} }

\def\endtitlepage{\if@restonecol\twocolumn \else \newpage \fi
        \def\thefootnote{\arabic{footnote}}
        \setcounter{footnote}{0}}  

\catcode`@=12 \relax

\def\figcap{\section*{Figure Captions\markboth
        {FIGURECAPTIONS}{FIGURECAPTIONS}}\list
        {Figure \arabic{enumi}:\hfill}{\settowidth\labelwidth{Figure
999:}
        \leftmargin\labelwidth
        \advance\leftmargin\labelsep\usecounter{enumi}}}
 \relax
\def\tablecap{\section*{Table Captions\markboth
        {TABLECAPTIONS}{TABLECAPTIONS}}\list
        {Table \arabic{enumi}:\hfill}{\settowidth\labelwidth{Table
999:}
        \leftmargin\labelwidth
        \advance\leftmargin\labelsep\usecounter{enumi}}}
 \relax
\def\reflist{\section*{References\markboth
        {REFLIST}{REFLIST}}\list
        {[\arabic{enumi}]\hfill}{\settowidth\labelwidth{[999]}
        \leftmargin\labelwidth
        \advance\leftmargin\labelsep\usecounter{enumi}}}
 \relax
\makeatletter
\newcounter{pubctr}
\def\publist{\@ifnextchar[{\@publist}{\@@publist}}
\def\@publist[#1]{\list
        {[\arabic{pubctr}]\hfill}{\settowidth\labelwidth{[999]}
        \leftmargin\labelwidth
        \advance\leftmargin\labelsep
        \@nmbrlisttrue\def\@listctr{pubctr}
        \setcounter{pubctr}{#1}\addtocounter{pubctr}{-1}}}
\def\@@publist{\list
        {[\arabic{pubctr}]\hfill}{\settowidth\labelwidth{[999]}
        \leftmargin\labelwidth
        \advance\leftmargin\labelsep
        \@nmbrlisttrue\def\@listctr{pubctr}}}
 \relax
\makeatother

\def\be{\begin{equation}}
\def\ee{\end{equation}}
\def\ba{\begin{eqnarray}}
\def\ea{\end{eqnarray}}

\def\m{\mu}

\def\l{\lambda}

\def\no{\noindent}

\def\IR{\relax{\rm I\kern-.18em R}}

\def\bse{\begin{small}\begin{equation*}}
\def\ese{\end{equation*}\end{small}}

\begin{document}


\renewcommand{\theequation}{\thesection.\arabic{equation}}
\csname @addtoreset\endcsname{equation}{section}

\newcommand{\eqn}[1]{(\ref{#1})}

\begin{titlepage}
\begin{center}
\strut\hfill
\vskip 1.3cm


\vskip .5in

{\Large \bf Classical impurities associated to high rank algebras}

\vskip 0.5in

{\large \bf Anastasia Doikou}\phantom{x}
\vskip 0.02in
{\footnotesize Department of Mathematics, Heriot-Watt University,\\
EH14 4AS, Edinburgh, United Kingdom}
\\[2mm]
\noindent
{\footnotesize and}

\vskip 0.02in
{\footnotesize Department of Computer Engineering \& Informatics,\\
 University of Patras, GR-Patras 26500, Greece}
\\[2mm]
\noindent
\vskip .1cm


{\footnotesize {\tt E-mail: A.Doikou@hw.ac.uk}}\\

\end{center}

\vskip 1.0in

\centerline{\bf Abstract}

Classical integrable impurities associated to high rank ($\mathfrak{gl}_N$) algebras are investigated. A particular prototype i.e. the vector non-linear Schr\"{o}dinger (NLS) model is chosen as an example. A systematic construction of local integrals of motion as well as the time components of the corresponding Lax pairs is presented based on the underlying classical algebra. Suitable gluing conditions compatible with integrability are also extracted. The defect contribution is also examined in the case where non-trivial integrable conditions are implemented. It turns out that the integrable boundaries may drastically alter the bulk behavior, and in particular the defect contribution.

\no

\vfill

\end{titlepage}
\vfill \eject

\tableofcontents

\section{Introduction}

Classical integrable field theories and quantum integrable lattice systems in the presence of point-like defects have been the subject of increasing research interest in recent years \cite{delmusi}--\cite{doikou-defects} with particular emphasis on the identification of the associated integrals of motion (conserved charges) as well as the derivation of the physical transmission matrices. The central purpose of the present article is the study of a classical vector NLS model in the presence of a point like defect. This model is associated to the $\mathfrak{gl}_N$ algebra, and the corresponding defect matrix is a generic representation of the aforementioned algebra.

This investigation is based on the underlying algebra as well as the corresponding linear auxiliary problem along the lines introduced in \cite{avan-doikou1}. Using these fundamental notions we are able to extract the associated integrals of motion, as well as the corresponding Lax pairs. Analyticity conditions on the time components of the Lax pair are also required in order to eliminate singular behavior in the zero curvature condition. This provides sewing conditions across the defect point that relate the left and right values of the fields and their derivatives. These are necessary conditions that allow identification of the Hamiltonian equations of motion.

A particularly interesting issue, which is also addressed in this study, is how the presence of non-trivial, but still integrable, conditions affects the bulk behavior of the system and more specifically the defect contribution. Two sets of distinct boundary conditions \cite{sklyanin} are introduced, i.e. the soliton preserving (SP) and the soliton non-preserving (SNP) associated to the reflection algebra and the twisted Yangian respectively. Based on the relevant boundary algebras we extract the first integrals of motion in each case, and we identify the corresponding defect contributions. Non-trivial bulk behavior is observed in particular in the case of the reflection algebra.

The outline of this paper is as follows: in the next section we review the general algebraic setting, which will be used in the present study.  In Section 3 we briefly review the vector NLS model, we recall the Lax pair as well as the construction of the first integrals of motion. In section 4 the model is investigated in the presence of a point-like defect. Explicit expressions of the first local integrals of motion, as well as the corresponding time components of the Lax pair are extracted. Suitable analyticity conditions are imposed and sewing conditions across the defect point are derived. The equations of motion for the bulk theories as well as the defect point are also identified.
In section 5 non trivial integrable boundary conditions are implemented, and the defect contributions are extracted based on the respective algebras. In the last section a brief discussion of our findings is presented together with some possible future research directions.

\section{The general setting}

Before we present the main results of this investigation about the vector NLS model
in the presence of a point-like defect let us first briefly review some
of the basic notions that will be used below. The formulation will be adopted for the study of the
$\mathfrak{gl}_N$ NLS model with defect is based on the
solution of the so called auxiliary linear problem \cite{ft}. It
is therefore necessary to recall at least the basics regarding
this formulation. Let $\Psi$ be a solution of the following set of
equations
\ba
&&{\partial \Psi \over
\partial x} = {\mathbb U}(x,t, \lambda) \Psi \label{dif1}
\\ &&
{\partial  \Psi \over \partial t } = {\mathbb V}(x,t,\lambda) \Psi
\label{dif2}
\ea
with ${\mathbb U},\ {\mathbb V}$ being in general
$n \times n$ matrices with entries functions of complex valued
fields, their derivatives, and the spectral parameter $\lambda$. The
monodromy matrix from (\ref{dif1}) may be then written as:
\be
T(x,y,\lambda) = {\cal P} \exp \Big \{ \int_{y}^x {\mathbb
U}(x',t,\lambda)dx' \Big \}. \label{trans}
\ee
The fact that $T$ is
a solution of equation (\ref{dif1}) will be extensively used
subsequently for obtaining the relevant integrals of motion.
Compatibility conditions of the two differential equations
(\ref{dif1}), (\ref{dif2}) lead to the zero curvature condition
\be
\dot{{\mathbb U}} - {\mathbb V}' + \Big [{\mathbb U},\ {\mathbb V}
\Big ]=0, \label{zecu}
\ee
giving rise to the corresponding
classical equations of motion of the system under consideration.

There exists an alternative description, known as the  $r$ matrix
approach (Hamiltonian formulation). In this picture the underlying
classical algebra is manifest in analogy to the quantum case as
will become quite transparent later. Let us first recall this
method for a general classical integrable system on the full line.
The existence of the classical $r$-matrix \cite{sts}, satisfying the classical Yang-Baxter equation
\be
\Big [r_{12}(\lambda_1-\lambda_2),\
r_{13}(\lambda_1)+r_{23}(\lambda_2) \Big]+ \Big
[r_{13}(\lambda_1),\ r_{23}(\lambda_2) \Big] =0,
\ee
guarantees
the integrability of the classical system.
Indeed, consider that ${\mathbb U}$ satisfies the linear algebra:
\be
\Big \{ {\mathbb U}_1(\lambda,x),\ {\mathbb U}_2(\mu, y) \Big \} = \Big [r_{12}(\lambda-\mu),\ {\mathbb U}_1(\lambda, x)+{\mathbb U}_2(\mu, y) \Big ]\ \delta(x-y).
\ee
Then is is straightforward to show that the
operator $T(x,y,\lambda)$ satisfies the classical quadratic algebra
\be \Big
\{T_{1}(x,y,t,\lambda_1),\ T_{2}(x,y,t,\lambda_2) \Big \} =
\Big[r_{12}(\lambda_1-\lambda_2),\
T_1(x,y,t,\lambda_1)\ T_2(x,y,t,\lambda_2) \Big ]. \label{basic}
 \ee
Making use of the latter equation one may readily show for a
system on the full line:
\be \Big \{\ln tr\{T(x,y,\lambda_1)\},\ \ln
tr\{T(x,y, \lambda_2)\} \Big\}=0
\ee
i.e. the system is
integrable, and the charges in involution --local integrals of
motion-- may be obtained via the expansion of the object $\ln
tr\{T(x,y,\lambda)\}$, based essentially on the fact that $T$ also
satisfies (\ref{dif1}).

\section{The vector NLS theory}

We shall hereafter focus on the $\mathfrak{gl}_N$ NLS model. Consider the corresponding
classical $r$-matrix \cite{yang}
\be
r(\lambda) = {\kappa {\mathrm P} \over \lambda}
~~~~\mbox{where} ~~~~{\mathrm P}=\sum_{i,j=1}^{N} e_{ij} \otimes
e_{ji} \label{rr}
\ee
${\mathrm P}$ is the $\mathfrak{gl}_N$ permutation operator,
and $e_{ij}$ are $ N \times N$ matrices such that: $~(e_{ij})_{kl} = \delta_{ik} \delta_{jl}$.
The Lax pair for
the generalized NLS model is given by the following expressions
\cite{ft}:
\be
{\mathbb U} = {\mathbb U}_0 + \lambda
{\mathbb U}_1, ~~~{\mathbb V} = {\mathbb V}_0+\lambda{\mathbb V}_1
+\lambda^2 {\mathbb V}_2,
\ee
where
\ba
&&
{\mathbb U}_1 = {1\over 2i} (\sum_{i=1}^{N-1}e_{ii} -e_{NN}),
~~~~{\mathbb U}_0 =
\sqrt{\kappa} \sum_{i=1}^{N-1}(\bar \psi_i e_{iN} +\psi_i e_{Ni}) \cr
&& {\mathbb V}_0 = i\kappa \sum_{i,\ j=1}^{N-1}(\bar \psi_i \psi_j
e_{ij} -|\psi_i|^2 e_{NN}) -i\sqrt{\kappa}\sum_{i=1}^{N-1} (\bar
\psi_i' e_{iN} - \psi_i' e_{Ni}), \cr
&&
{\mathbb V}_1= -{\mathbb U}_{0}, ~~~{\mathbb V}_2= -{\mathbb U}_{1} \label{lax}
\ea
and $\psi_i,\ \bar \psi_j$ satisfy\footnote{The Poisson
structure for the generalized NLS model is defined as:
\be
\Big \{
A,\ B  \Big \}=  -i \sum_{i} \int_{-L}^{L} dx \Big ({\delta A \over
\delta \psi_i(x)}\ {\delta B \over \delta \bar \psi_i(x)} -
{\delta A \over \delta \bar \psi_i(x)}\ {\delta B \over \delta
\psi_i(x)}\Big )
\ee}:
\be
\Big \{ \psi_{i}(x),\ \psi_j(y) \Big
\} = \Big \{\bar \psi_{i}(x),\ \bar \psi_j(y) \Big \} =0, ~~~~\Big
\{\psi_{i}(x),\ \bar \psi_j(y) \Big \}= -i \delta_{ij}\ \delta(x-y).
\ee
From the zero curvature condition (\ref{zecu}) the classical
equations of motion for the generalized NLS model are entailed
i.e.
\be
i{\partial \psi_{i}(x,t) \over
\partial t} = - {\partial^{2} \psi_{i}(x,t) \over
\partial^2 x}+2\kappa \sum_{j}|\psi_{j}(x,t)|^2 \psi_{i}(x,t),
~~~~i,\ j \in \{1, \ldots, N-1 \}. \label{nls}
\ee
It is clear
that for $N=2$ the equations of motion of the usual NLS model are
recovered.

As already mentioned to obtain the local integrals of motion of the
NLS model one has to expand $T$ (\ref{trans}) in powers of
$\lambda^{-1}$ \cite{ft}. Let us consider the following ansatz for
$T$ as $|\lambda| \to \infty$
\be
T(x,y,\lambda) = ({\mathbb I}
+W(x, \lambda))\ \exp[Z(x,y,\lambda)]\ ({\mathbb I}
+W(y,\lambda))^{-1} \label{exp0}
\ee
where $W$ is off diagonal
matrix i.e. $~W = \sum_{i\neq j} W_{ij} e_{ij}$, and $Z$ is purely
diagonal $~Z = \sum_{i=1}^N Z_{ii}e_{ii}$.
Also
\be
Z_{ii}(\lambda) = \sum_{n=-1}^{\infty} {Z^{(n)}_{ii} \over
\lambda^{n}}, ~~~~W_{ij} = \sum_{n=1}^{\infty}{W_{ij}^{(n)} \over
\lambda^n}. \label{expa}
\ee
Inserting the latter expressions
(\ref{expa}) in (\ref{dif1}) one may identify the coefficients
$W_{ij}^{(n)}$ and $Z_{ii}^{(n)}$ (see also \cite{ft, doikou-fioravanti-ravanini}). Notice that as $i\lambda \to \infty$ the only non
negligible contribution from $Z^{(n)}$ comes from the
$Z^{(n)}_{NN}$ term, and is given by:
\be Z_{NN}^{(n)}(L, -L)= i L
\delta_{n, -1} +\sqrt{\kappa}\sum_{i=1}^{N-1}\int_{-L}^L dx\
\psi_i(x)\ W_{iN}^{(n)}(x). \label{ref1}
\ee
It is thus sufficient
to determine the coefficients $W_{iN}^{(n)}$ in order to extract
the relevant local integrals of motion.
Indeed solving (\ref{dif1}) one may easily obtain \cite{doikou-fioravanti-ravanini}:
\ba &&
W_{iN}^{(1)}(x) = -i \sqrt{\kappa} \bar \psi_{i}(x),
~~~~W_{iN}^{(2)}(x)=\sqrt{\kappa} \bar \psi_i'(x)\cr
&&
W_{iN}^{(3)}(x) = i\sqrt{\kappa}\bar \psi_i''(x) -i
\kappa^{{3\over 2}} \sum_{k} |\psi_{k}(x)|^2\bar \psi_i(x),
~~\ldots. \label{ref2}
\ea
From the latter formulae (\ref{ref2})
and taking into account (\ref{exp0}), (\ref{ref1}) the local
integrals of motion of NLS may be readily extracted from $\ln\ tr
T(L, -L, \lambda)$, i.e.
\ba && I_1 = -i\kappa \int_{-L}^{L} dx\
\sum_{i=1}^{N-1}\psi_i(x) \bar \psi_i(x) , \cr
&& I_2 =-{\kappa
\over 2} \int_{-L}^{L} dx\ \sum_{i=1}^{N-1} \Big (\bar \psi_i(x)
\psi'_i(x)- \psi_i(x) \bar \psi'_i(x) \Big ),
\cr
&& I_3=
-i\kappa \int_{-L}^{L} dx\ \sum_{i=1}^{N-1} \Big (\kappa
|\psi_i(x)|^2 \sum_k|\psi_k(x)|^2  +\psi_i'(x) \bar \psi_i'(x)
\Big ). \label{first}
\ea
The corresponding familiar quantities for the generalized NLS are given by:
\be {\cal
N}=-{I_1 \over i \kappa},~~~~{\cal P}=-{I_2 \over i \kappa},
~~~~{\cal H}=-{I_3 \over i \kappa},
\label{clas}
\ee
and apparently
\be \{{\cal H},\ {\cal P}\}=\{{\cal H},\ {\cal N}\}=\{ {\cal N},\
{\cal P} \}=0.
\ee
Again in the special case where $N=2$ the well known local integrals of motion for the usual NLS model on the
full line are recovered.

In the
case of integrable boundary conditions, which will be treated later in the text,
we shall need in
addition to (\ref{ref2}) the following objects:
\ba
&&
W_{Ni}^{(1)} = i\sqrt{\kappa}\psi_i, ~~~~W_{Ni}^{(2)} = -i
W^{'(1)}_{Ni} +\sum_{i\neq j,\ i\neq N,\ j\neq N}W^{(1)}_{N
j}W_{ji}^{(1)},~~~~W_{ji}^{'(1)} = iW_{jN}^{(1)} W_{Ni}^{(1)}
\cr
&& W_{Ni}^{(3)} = - iW^{'(2)}_{Ni}
+W_{iN}^{(1)}W_{Ni}^{(1)}W_{Ni}^{(1)} + \sum_{i\neq j,\ i \neq N,
j\neq N} W_{Nj}^{(1)}W_{ji}^{(2)} \cr
&& W_{ij}^{'(2)} =
iW_{iN}^{(1)} W_{Nj}^{(2)}-i W_{jN}^{(1)} W_{Nj}^{(1)}
W_{ij}^{(1)}.
\label{reff}\ea
We shall also need for our computations here the following:
\ba
&& (1+W(\lambda))^{-1} = 1 + F(\lambda)
~~~\mbox{where} ~~~F(\lambda) = \sum_{n=1}^{\infty} {{\mathrm
f}^{(n)} \over \lambda^n}, \label{nots0} \cr
&& \mbox{where} ~~~~~ {\mathrm f}^{(1)}
= -W^{(1)}, ~~~~{\mathrm f}^{(2)} = (W^{(1)})^2 -W^{(2)},\cr
&& {\mathrm f}^{(3)} = -W^{(3)} + W^{(1)}W^{(2)} + W^{(2)} W^{(1)} - (W^{(1)})^3,
~~~\ldots \label{nots}
\ea

\section{The vector NLS model in the presence of defect}

In this section the local integrals of motion as well as the associated Lax pairs for the NLS model in the presence of a point like integrable defect at $x=x_0$ will be derived. To achieve this we first need to introduce the key element in our study, which is the modified monodromy matrix \cite{avan-doikou1}
\be
T(\lambda) = T^{+}(L, x_0;\lambda)\ {\mathbb L}(x_0; \lambda)\ T^-(x_0, -L; \lambda), \label{mono1}
\ee
where we define
\ba
&& T^{+}(L,x_0; \lambda) = P \exp\Big [ \int_{x_0}^L dx\ {\mathbb U}^{+}(x)\Big], ~~~~T^{-}(x_0, -L; \lambda) = P \exp\Big [ \int^{x_0}_{-L} dx\ {\mathbb U}^{-}(x)\Big], \cr
&& {\mathbb L}(x_0; \lambda) = \lambda + \kappa {\mathbb P}(x_0), ~~~~~~{\mathbb P} = \sum_{i,j} {\mathbb P}_{ij}\ e_{ij}.
\ea
The matrices ${\mathbb U}^{\pm}$ together with the corresponding time components ${\mathbb V}^{\pm}$ satisfy the
familiar linear auxiliary problem and they naturally satisfy the zero curvature condition:
\ba
&&{\partial \Psi \over
\partial x} = {\mathbb U}^{\pm}(x, t, \lambda) \Psi \label{dif1d}
\\ &&
{\partial  \Psi \over \partial t } = {\mathbb V}^{\pm}(x,t,\lambda) \Psi, ~~~~~x\neq x_0.
\label{dif2d}
\ea
Compatibility conditions of the two differential equations
(\ref{dif1d}), (\ref{dif2d}) lead to the zero curvature condition
\be
\dot{{\mathbb U}}^{\pm} - {\mathbb V}^{\pm'} + \Big [{\mathbb U}^{\pm},\ {\mathbb V}^{\pm}
\Big ]=0, ~~~~x \neq x_0,\label{zecud}
\ee
giving rise to the corresponding
classical equations of motion for the corresponding bulk theories. However, special care should be taken regarding the defect point. On the defect point the zero curvature condition is quite modified (see relevant discussion, e.g. in \cite{avan-doikou1}) and takes the form:
\be \dot {\mathbb L}(\lambda) = \tilde{\mathbb V}^{+}(\lambda)\ {\mathbb L}(\lambda) - {\mathbb L}(\lambda)\ \tilde {\mathbb V}^-(\lambda). \label{zero-defect}
\ee
The latter naturally leads to the equations of motion on the defect point. $\tilde{\mathbb V}^{\pm}$ are the time components of the Lax pair on the defect point from the left and from the right (see also \cite{avan-doikou1}).

The matrices ${\mathbb U}^{\pm}$ satisfy the ultra-local linear algebra:
\be
\Big \{{\mathbb U}^{\pm}_1(x, \lambda),\ {\mathbb U}^{\pm}_2(y, \mu) \Big \} = \Big [r_{12}(\lambda -\mu),\ {\mathbb U}^{\pm}_1(x, \lambda)+ {\mathbb U}^{\pm}_2(y, \mu) \Big ]\ \delta(x-y),
\ee
giving rise to the following exchange relations:
\be
\Big \{ \psi^{\pm}_i(x),\ \psi^{\pm}_j(y) \Big \} =0, ~~~\Big \{ \bar \psi^{\pm}_i(x),\ \bar \psi^{\pm}_j(y) \Big \} =0, ~~~\Big \{ \psi^{\pm}_i(x),\ \bar \psi^{\pm}_j(y) \Big \} = -i\delta_{ij}\ \delta(x-y).
\ee
The defect ${\mathbb L}$-matrix satisfies the same quadratic algebra with the bulk monodromy matrices
$T^{\pm}$ i.e.
\be
\Big \{{\mathbb L}_1(\lambda),\ {\mathbb L}_2(\mu) \Big \} = \Big [r_{12}(\lambda - \mu),\ {\mathbb L}_1(\lambda)\ {\mathbb L}_2(\mu) \Big ]. \label{rttbb}
\ee
In fact, this is the key requirement that ensures the integrability of the model \cite{avan-doikou1}.
It is clear that due to the quadratic algebraic relation (\ref{rttbb}) the following exchange relations regarding the defect degrees of freedom arise
\be
\Big \{{\mathbb P}_{ij},\ {\mathbb P}_{mn} \Big \} = {\mathbb P}_{in} \delta_{jm} - {\mathbb P}_{mj} \delta_{in}, \label{gln}
\ee
which are the exchange relations of the classical $\mathfrak{gl}_N$ algebra (see e.g. \cite{ft}).

\subsection{The local integrals of motion}

The local integrals of motion ${\cal I}_m$  will be obtained subsequently via the familiar generating function, i.e. the corresponding transfer matrix, --the trace of the modified monodromy matrix--
\be
\ln t(\lambda) = \sum_m {{\cal I}_m \over \lambda^m}.
\ee
Taking into account the discussion of section 3 we conclude:
\be
\ln t(\lambda) = Z_{NN}^{+}(\lambda) +Z^-_{NN}(\lambda) + \ln \Big [ (1 +W^+(x_0))^{-1}\ {\mathbb L}(x_0)\ (1+W^-(x_0))\Big ]_{NN}.
\ee
The last term of the latter expression provides the defect contribution, whereas clearly the first two terms give the left and right bulk theory contributions.

Taking also into account the information provided in section 3, and carefully expanding the generating function we conclude that:
\be
{\cal I}_1 = -i\kappa \int_{-L}^{x_0} dx\
\sum_{i=1}^{N-1}\psi^-_i(x) \bar \psi^-_i(x)  -i\kappa \int_{x_0}^{L} dx\
\sum_{i=1}^{N-1}\psi^+_i(x) \bar \psi^+_i(x)+ {\cal D}_1,
\ee
\ba
&& {\cal I}_2 =-{\kappa
\over 2} \int_{-L}^{x_0} dx\ \sum_{i=1}^{N-1} \Big (\bar \psi^-_i(x)
\psi^{-'}_i(x)- \psi^-_i(x) \bar \psi^{-'}_i(x) \Big ) +{\kappa \over 2}\sum_i\psi_i^-(x_0)\bar \psi_i^-(x_0)
\cr
&& -{\kappa
\over 2} \int_{x_0}^{L} dx\ \sum_{i=1}^{N-1} \Big (\bar \psi^+_i(x)
\psi^{+'}_i(x)- \psi_i^+(x) \bar \psi^{'+}_i(x) \Big ) - {\kappa \over 2}\sum_i \psi_i^+(x_0)\bar \psi_i^+(x_0) +{\cal D}_2 \cr &&
\ea
\ba
&& {\cal I}_3=
-i\kappa \int_{-L}^{x_0} dx\ \sum_{i=1}^{N-1} \Big (\kappa
|\psi^-_i(x)|^2 \sum_k|\psi^-_k(x)|^2  +\psi_i^{-'}(x) \bar \psi_i^{-'}(x)
\Big ) + i \kappa \sum_i \psi_i^-(x_0) \bar \psi_i^{-'}(x_0)  \cr
&& -i\kappa \int_{x_0}^{L} dx\ \sum_{i=1}^{N-1} \Big (\kappa
|\psi^+_i(x)|^2 \sum_k|\psi^+_k(x)|^2  +\psi_i^{+'}(x) \bar \psi_i^{+'}(x)
\Big )-  i \kappa \sum_i \psi_i^+(x_0) \bar \psi_i^{+'}(x_0)   + {\cal D}_3, \cr
&& \label{first1}
\ea
where the defect contributions are defined as
\ba
&& {\cal D}_1 = D_1 \cr
&& {\cal D}_2 = D_2 - {D_1^2 \over 2}\cr
&& {\cal D}_3 = D_3 - D_1 D_2 +{D_1^3 \over 3}
\ea
and:
\ba
&& D_1 = \kappa {\mathbb P}_{NN} \cr
&& D_2 =  \sum_{j\neq N} W_{Nj}^{+(1)}W_{jN}^{+(1)} - \kappa \sum_{j\neq N} W_{Nj}^{+(1)}{\mathbb P}_{jN} +\kappa \sum_{j\neq N} {\mathbb P}_{Nj} W_{jN}^{-(1)}- \sum_{j\neq N} W_{Nj}^{+(1)}W_{jN}^{-(1)}\cr
&& D_3 = \sum_{j \neq N} W_{Nj}^{+(1)}W_{jN}^{+(2)}+\sum_{j \neq N} W_{Nj}^{+(2)}W_{jN}^{+(1)} - \sum_{i\neq N  i, j \neq N} W_{Ni}^{+(1)} W_{ij}^{+(1)}W_{jN}^{+(1)}  - \sum_{j \neq N} W_{Nj}^{+(1)}W_{jN}^{-(2)} \cr
&&+ \kappa \sum_{j \neq N} {\mathbb P}_{Nj} W_{jN}^{-(2)}
+ \sum_{i\neq N, j \neq N} W_{Ni}^{+(1)} W_{ij}^{+(1)}W_{jN}^{-(1)} - \sum_{j\neq N} W_{Nj}^{+(2)}W_{jN}^{-(1)} -\sum_{i\neq N,j\neq N} W_{Nj}^{+(1)}{\mathbb P}_{ij} W_{jN}^{-(1)}. \label{dd} \cr
&& -\kappa \sum_{j\neq N}W_{Nj}^{+(2)}{\mathbb P}_{jN} + \kappa \sum_{i\neq N,\ j}W_{Ni}^{+(1)} W_{ij}^{+(1)} {\mathbb P}_{jN}
\ea
$W^{\pm(n)}_{kl}$ are expressed in terms of the fields $\psi^{\pm},\ \bar \psi^{\pm}$ and their derivatives as in (\ref{ref2}).

As in the periodic case reviewed in section 3, the first three integrals of motion are respectively: the numbers of particles, the momentum and the Hamiltonian:
\be
{\cal N} = -{{\cal I}_1 \over i \kappa},~~~~{\cal P}=-{{\cal I}_2 \over i \kappa},
~~~~{\cal H}=-{{\cal I}_3 \over i \kappa},
\label{clas1}
\ee
and apparently
\be
\{{\cal H},\ {\cal P}\}=\{{\cal H},\ {\cal N}\}=\{ {\cal N},\
{\cal P} \}=0.
\ee

The results of this subsection are clearly reduced to the ones of \cite{avan-doikou1} in the case where $N=2$.
To this point no gluing conditions have been determined; to achieve this we shall explicitly derive the Lax pair time component associated to each integral of motion, and then impose suitable analyticity conditions (see also \cite{avan-doikou1}). This will be the subject of the subsequent subsection.

\subsection{The Lax pair}

Based on the underlying classical algebra one can extract the time component ${\mathbb V}$ of the Lax pair.
In the case where defects are taken into account, one has to compute $\mathbb{V}$ in
the bulk, as well as at the defect point. If the $r$-matrix of
the model is the Yangian, as it happens in the vector NLS model, the corresponding time components
are simplified and for a single point-like defect are expressed as \cite{avan-doikou1}
\ba
&& \mathbb{V}^+(x,\l,\m) = \frac{t^{-1}(\lambda)}{\l-\m}T^+(x,x_0){\mathbb L}(x_0)
T^-(x_0,-L)T^+(L,x) \cr
&& \mathbb{V}^-(x,\l,\m) = \frac{t^{-1}(\lambda)}{\l-\m}T^-(x,-L)T^+(L,x_0) {\mathbb L}(x_0)
T^-(x_0,x)\cr
&& \mathbb{\widetilde{V}}^+(x_0,\l,\m) = \frac{t^{-1}(\lambda)}{\l-\m} {\mathbb L}(x_0)T^-(x_0,-L)
T^+(L,x_0) \cr
&& \mathbb{\widetilde{V}}^-(x_0,\l,\m) = \frac{t^{-1}(\lambda)}{\l-\m} T^-(x_0,-L)T^+(L,x_0)
{\mathbb L}(x_0),
\ea
where ${\mathbb V}^{\pm}$ are the bulk left and right quantities, and $\mathbb{\widetilde{V}}^{\pm}$ are the quantities associated to the defect point from the left and the right. Substituting the expressions of $T^{\pm}$ in the formulas above we obtain the usual ${\mathbb V}^{\pm (i)}$ bulk matrices, and the defect point Lax pair time components:
\ba
&& \tilde {\mathbb V}^{- (1)} = e_{NN} \cr
&& \tilde {\mathbb V}^{-(2)}(\lambda) = \lambda\ e_{NN} - \sum_{j\neq N} (W_{Nj}^{+(1)} - \kappa {\mathbb P}_{Nj})\ e_{Nj} + \sum_{j\neq N}W_{jN}^{-(1)}\ e_{jN} \cr
&& \tilde {\mathbb V}^{-(3)}(\lambda) =  \lambda\ \tilde {\mathbb V}^{-(2)}(\lambda) + {\mathfrak V}^-
\ea
and we define,
\ba
&& {\mathfrak V}^- =  \sum_{j\neq N} \Big (W_{Nj}^{+(1)}W_{jN}^{-(1)} - \kappa
{\mathbb P}_{Nj} W_{jN}^{-(1)} \Big )\ e_{NN} - \sum_{i \neq N, j \neq N}\Big ( -\kappa {\mathbb P}_{Nj} W_{iN}^{-(1)}+  W_{Nj}^{+(1)}W_{iN}^{-(1)}\Big )\ e_{ij} \cr
&& + \sum_{j\neq N}\Big ( -\kappa
\sum_{i\neq N} W_{Ni}^{+(1)} {\mathbb P}_{ij} + i W_{Nj}^{+(1)'}-\kappa^2 {\mathbb P}_{NN} {\mathbb P}_{Nj}+ \kappa {\mathbb P}_{NN} W_{Nj}^{+(1)}\Big )\ e_{Nj} +i\sum_{j\neq N} W_{jN}^{-(1)'} e_{jN}. \cr && \label{B-}
\ea
Similarly, for the $\tilde {\mathbb V}^+$ we find the following expressions:
\ba
&& \tilde {\mathbb V}^{+(1)}= e_{NN} \cr
&& \tilde {\mathbb V}^{+(2)}(\lambda) =  \lambda\ e_{NN} + \sum_{j\neq N} (\kappa {\mathbb P}_{jN} + W_{jN}^{-(1)})\ e_{jN} -\sum_{j\ne N} W_{Nj}^{+(1)}\ e_{Nj}\cr &&
\tilde {\mathbb V}^{+(3)}(\lambda)= \lambda \tilde  {\mathbb V}^{+(2)}(\lambda) + {\mathfrak V}^+
\ea
where
\ba
&&{\mathfrak V}^+ = \Big( \sum_{j\neq N} W_{Nj}^{+(1)} W_{jN}^{-(1)} + \kappa \sum_{j \neq j}W_{Nj}^{+(1)} {\mathbb P}_{jN} \Big ) e_{NN} -\sum_{i \neq N, j \neq N} \Big (W_{Nj}^{+(1)}W_{iN}^{-(1)} + \kappa W_{Nj}^{+(1)}{\mathbb P}_{iN} \Big ) e_{ij}\cr
&& + i \sum_{j\neq N} W_{Nj}^{+(1)'}e_{Nj} + \sum_{j\neq N} \Big ( \kappa\sum_{i\neq N} {\mathbb P}_{ji} W_{iN}^{-(1)}+ i W_{jN}^{-(1)'} - \kappa^2 {\mathbb P}_{NN} {\mathbb P}_{jN} - \kappa {\mathbb P}_{NN} W^{-(1)}_{jN} \Big )e_{jN}. \cr
&& \label{B+}
\ea

As explained in detail in \cite{avan-doikou1} analyticity conditions around the defect point lead to:
\be
\mathbb{V}^{\pm}(x_0^{\pm}) \to \mathbb{\widetilde{V}}^{\pm}(x_0),
\ee
which in turn give rise to the related sewing conditions. Indeed, the second order sewing conditions are given as
\ba
&& W_{Nj}^{+(1)}- W_{Nj}^{-(1)} = \kappa {\mathbb P}_{Nj} \cr
&& W_{jN}^{+(1)} - W_{jN}^{-(1)} =  \kappa {\mathbb P}_{jN},
 \label{sew1}
\ea
whereas the third order sewing conditions from the left and from the right are:
\ba
&& W_{Nj}^{+(1)'} - W_{Nj}^{-(1)'} =  i \kappa  \Big (-\sum_{i\neq N} W_{Ni}^{+(1)} {\mathbb P}_{ij} -\kappa {\mathbb P}_{NN} {\mathbb P}_{Nj}+ {\mathbb P}_{NN}W_{Nj}^{+(1)} \Big ) \cr
&& W_{jN}^{+(1)'} - W_{jN}^{-(1)'} = i\kappa \Big  (- \sum_{i\neq N} {\mathbb P}_{ji} W_{iN}^{-(1)} + \kappa {\mathbb P}_{jN} {\mathbb P}_{NN} + W_{jN}^{-(1)} {\mathbb P}_{NN} \Big  ). \label{sew2}
\ea
It is important to note that the sewing conditions above are compatible with the Hamiltonian action as has already been proven in \cite{avan-doikou1} for any system associated to the classical Yangian $r$-matrix.

Next we shall derive the equations of motion of the model under study in the bulk a well as on the defect point. These equations may be extracted from the Hamiltonian or the zero curvature conditions in the bulk and on the defect point (\ref{zero-defect}), and provide the time evolution of the dynamical degrees of freedom associated to the defect.
The equations of motion in the bulk are the familiar equations of the vector NLS model.

Let us now focus on the derivation of the equations of motion on the defect point.
In this case in order to cancel out the $\lambda$ dependence in (\ref{zero-defect}), we make use of the sewing conditions (\ref{sew1}), (\ref{sew2}). The contribution of the $\lambda$-independent part of the zero curvature condition on the defect point leads to the following set of equations, expressed in a compact form:
\be
\dot {\mathbb P}_{ij}= \sum_{l=1}^N{\mathfrak V}^+_{il}\ {\mathbb P}_{lj} - \sum_{l=1}^N{\mathbb P}_{il}\ {\mathfrak V}_{lj}^-, ~~~~~i,\ j \in \{1, \ldots, N\}.
\ee
In general, we define:
\be
{\mathfrak V}^{\pm} =\sum_{k, l} {\mathfrak V}^{\pm}_{kl}\ e_{kl},
\ee
and the elements ${\mathfrak V}^{\pm}_{kl}$ are defined in (\ref{B-}), (\ref{B+}).
Note that for the moment no sewing conditions have been implemented in the latter equations. Similar sets of equations associated to the defect have been extracted for a variety of integrable models (see e.g. \cite{avan-doikou1, avan-doikou, doikou-karaiskos-sigma}), thus the natural next step is to solve these equations using apparently the associated sewing conditions. This is an intriguing issue, which will be hopefully addressed in future investigations. In the special case where $N=2$ the results as expected reduce to the ones derived in \cite{avan-doikou1}.

\section{Defects in the presence of non-trivial boundaries}

In this section we shall examine the behavior of the point-like defect in the presence of non-trivial integrable boundaries. For this purpose Sklyanin's modified monodromy matrix \cite{sklyanin} will be considered to suitably incorporate the boundary effects. We shall also distinguish two types of boundary conditions the soliton non-preserving (SNP) associated to the so-called twisted Yangian and the soliton preserving (SP) ones associated to the reflection algebra.

\subsection{The SNP case}

This case is associated to the classical twisted Yangian \cite{sklyanin} defined as
\ba
 \Big \{ {\cal T}_1(\lambda_1),\ {\cal T}_2(\lambda_2) \Big\} &=& \Big [r_{12}(\lambda_1-\lambda_2),\ {\cal T}_1(\lambda_1)\ {\cal T}_2(\lambda_2)  \Big ] \cr
&+& {\cal T}_1(\lambda_1)r_{12}^{t_1}(-\lambda_1-\lambda_2){\cal T}_2(\lambda_2) - {\cal T}_2(\lambda_2) r_{12}^{t_1}(-\lambda_1 -\lambda_2){\cal T}_1(\lambda_1). \cr && \label{TY}
\ea
The modified monodromy matrix, representation of the classical quadratic algebra (\ref{TY}), according to Sklyanin \cite{sklyanin} is:
\be
{\cal T} = T(\lambda)\ K^-(\lambda)\ T^{t}(-\lambda), \label{rept1}
\ee
where the bulk monodromy matrix $T$ contains the point like defect, and is given by (\ref{mono1}). The generating function of the integrals of motion then becomes
\be
{\cal G}(\lambda) = \ln\ tr (K^+(\lambda)\ {\cal T}(\lambda)) = \sum_m {{\cal I}_m \over \lambda^m}. \label{bound1}
\ee
The $K^{\pm}$-matrices are $c$-number matrices, satisfying the respective algebra, with
\be
\Big \{K^{\pm}_1(\lambda_1),\ K_2^{\pm}(\lambda_2) \Big \} =0.
\ee
In what follows we shall consider for simplicity, but without loss of generality, $K^{\pm} \propto {\mathbb I}$.

It is technically more tractable to study the vector NLS model first and then take the suitable continuum limit (see also e.g. \cite{doikou-fioravanti-ravanini}). Let us briefly review the model.
The discrete monodromy matrix in the presence of a single defect on the $n$-th cite:
\be
T_0(L, 1; \lambda) = {\mathbb L}_{0L}(\lambda)\ldots \tilde {\mathbb L}_{0n}(\lambda) \ldots {\mathbb L}_{01}(\lambda)
\ee
Both ${\mathbb L}$ and $\tilde {\mathbb L}$, and consequently $T$ satisfy the classical quadratic algebra (\ref{basic}).
Where the ${\mathbb L}$ matrix of the discrete vector NLS model is given as
\be
{\mathbb L}(\lambda) = (i \lambda + \kappa {\mathbb N} )\ e_{NN} + \sum_{l = 1}^{N-1}e_{ll} + \sum_{l=1}^{N-1}(\phi_{l}\ e_{lN} + \psi_{l}\ e_{Nl}).
\ee
Due to the fact that ${\mathbb L}$ satisfies the quadratic algebra we obtain:
\be
\Big \{\phi_{k},\ \psi_{l} \Big \} = i \delta_{kl}.
\ee
The defect Lax operator is
\be
\tilde {\mathbb L}_n(\lambda) = \lambda + \kappa {\mathbb P}^{(n)}.
\ee
the entries of the ${\mathbb P}$ matrix ${\mathbb P}_{ij}$ satisfy the classical $\mathfrak{gl}_N$ algebra (\ref{gln}).

As already mentioned we focus  on the SNP case that is we consider the representation of the classical twisted Yangian (\ref{TY}). Expansion of $tr {\cal T}(\lambda)$, --${\cal T}$ defined in (\ref{rept1}), but now $T$ is the discrete monodromy matrix-- will lead to the local integrals of motion (see also \cite{doikou-fioravanti-ravanini} for relevant results). In this case only the {\it even} integrals of motion survive, the first non-trivial integral of motion turns out to be the discrete momentum:
\ba
{\cal I}^{(2)}_d &=& \kappa \sum_{l=1}^{N-1} \sum_{j \neq n, n-1} \psi_l^{(j+1)} \phi_l^{(j)} + i \kappa \sqrt{\kappa} \sum_{l=1}^{N-1}\Big ( \psi_l^{(n+1)} {\mathbb P}_{lN}^{(n)} + {\mathbb P}_{Nl}^{(n)} \phi^{(n-1)}_l \Big ) - \kappa^2 \sum_{j=1}^{N-1}({\mathbb N}^{(j)})^2 \cr
&+& \kappa \sum_{l=1}^{N-1} \psi_l^{(n+1)} \phi_l^{(n-1)} - \kappa \sum_{l=1}^{N-1}\Big (\phi_l^{(L)}\phi_l^{(L)} + \psi_l^{(1)}\psi_l^{(1)} \Big ).
\ea
By taking the continuum limit (see also \cite{avan-doikou1}, and references therein)
\ba
&& \Big ( \psi_l^{(j)},\ \phi^{(j)}_l \Big ) \to \Big ( \psi^+_l(x),\ \bar \psi^+_l(x) \Big ), ~~~~~~j > n, ~~~~x > x_0 \cr
&& \Big ( \psi_l^{(j)},\ \phi^{(j)}_l \Big ) \to \Big ( \psi^-_l(x),\ \bar \psi^-_l(x) \Big ), ~~~~~~j < n, ~~~~x < x_0 \cr
&& f^{(j+1)} \to f(x+\delta) \label{climit}
\ea
we obtain
    \ba
   && {\cal I}_2 =- \kappa \int_{0}^{x_0} dx\ \sum_{i=1}^{N-1} \Big (\bar \psi^-_i(x)
\psi^{-'}_i(x)- \psi^-_i(x) \bar \psi^{-'}_i(x) \Big ) +\kappa \sum_i\psi_i^-(x_0)\bar \psi_i^-(x_0)
\cr
&& -\kappa
 \int_{x_0}^{L} dx\ \sum_{i=1}^{N-1} \Big (\bar \psi^+_i(x)
\psi^{+'}_i(x)- \psi_i^+(x) \bar \psi^{'+}_i(x) \Big ) - \kappa \sum_i \psi_i^+(x_0)\bar \psi_i^+(x_0)  \cr && +\kappa \sum_i\psi_i^-(0)\bar \psi_i^-(0) +  \kappa \sum_i \psi_i^-(0) \psi_i^-(0) +2 {\cal D}_2  \cr &&
    \ea
   ${\cal D}_2$ is defined in (\ref{dd}).
    The charge ${\cal I}_2$ gives the momentum of the system:
    \be
    {\cal P} = - { {\cal I}_2 \over 2 i \kappa }.
    \ee
Note that in the SNP case although only the even charges are conserved the form of the momentum for instance is not drastically altered compared to the periodic case studied in the previous section. This however does not hold in the SP case, where significant modifications in the defect behavior are manifest as will be transparent subsequently.

\subsection{The SP case}

The SP case is associated to the classical reflection algebra \cite{sklyanin}
\ba
\Big \{ {\cal T}_1(\lambda_1),\ {\cal T}_2(\lambda_2) \Big\} &=& \Big [r_{12}(\lambda_1-\lambda_2),\ {\cal T}_1(\lambda_1)\ {\cal T}_2(\lambda_2)  \Big ] \cr
&+& {\cal T}_1(\lambda_1)r_{12}(\lambda_1+\lambda_2){\cal T}_2(\lambda_2) - {\cal T}_2(\lambda_2) r_{12}(\lambda_1 +\lambda_2){\cal T}_1(\lambda_1). \cr &&
\ea
The generic representation of the reflection algebra \cite{sklyanin}
\be
{\cal T} = T(\lambda)\ K^-(\lambda)\ T^{-1}(-\lambda).
\ee
The generating function of the integrals of motion in the SP case is then expressed as (\ref{bound1}), where now ${\cal T}$ is given above.

As in the SNP cases the $K^{\pm}$-matrices are $c$-number matrices, satisfying the respective algebra,
We shall also consider here for simplicity, $K^{\pm} \propto {\mathbb I}$. As will be clear below the particular choice of $K^{\pm}$ does not affect the defect contribution, thus it is convenient to consider the simplest $K$-matrices. We shall also assume for brevity Schwartz boundary conditions at $x= L$, i.e. the fields and their derivatives are zero. For generic boundary conditions especially in the SP case see \cite{doikou-fioravanti-ravanini}.

In this case the {\it odd} charges survive, we provide below the first non-trivial charges i.e. the number of particles and the Hamiltonian. We expand the generating function in powers of ${1\over \lambda}$ to obtain the relevant charges in involution. The leading contribution comes from the $Z_{NN},\ \hat Z_{NN}$ terms, (see also relevant discussion in section 3), thus:
\ba
\ln\ tr(K^+(\lambda)\ {\cal T}(\lambda)) &=& Z_{NN}^+(L,x_0) + Z_{NN}^-(x_0, 0) - \hat Z_{NN}^+(L, x_0) -\hat Z_{NN}^-(x_0, 0)\cr
&+& \ln \Big [ (1+W^-(0))^{-1}K^-(\lambda)(1+\hat W^{-}(0))\Big ]_{NN} \cr
&+& \ln \Big [ (1+ \hat W^+(L))^{-1}K^+(\lambda)(1 + W^{-}(L))\Big ]_{NN}\cr
&+&\ln \Big [ (1+W^+(x_0))^{-1}  {\mathbb L}(x_0)(1+W^-(x_0))\Big ]_{NN} \cr &+& \ln \Big [ (1+ \hat W^-(x_0))^{-1} \hat {\mathbb L}^{-1}(x_0)(1+\hat W^+(x_0)) \Big ]_{NN}. \label{bound2}
\ea
we introduce the notation: $\hat f(\lambda) = f(-\lambda)$.
The associated conserved quantities are then given as:
\be
{\cal I}_1 = -2i\kappa \int_{0}^{x_0} dx\
\sum_{i=1}^{N-1}\psi^-_i(x) \bar \psi^-_i(x)  -2i\kappa \int_{x_0}^{L} dx\
\sum_{i=1}^{N-1}\psi^+_i(x) \bar \psi^+_i(x)+ 2 {\cal D}_1,
\ee
\ba
&& {\cal I}_3=
-2i\kappa \int_{0}^{x_0} dx\ \sum_{i=1}^{N-1} \Big (\kappa
|\psi^-_i(x)|^2 \sum_k|\psi^-_k(x)|^2  +\psi_i^{-'}(x) \bar \psi_i^{-'}(x)
\Big ) + 2i \kappa \sum_i \psi_i^-(x_0) \bar \psi_i^{-'}(x_0) \cr
&& -2i\kappa \int_{x_0}^{L} dx\ \sum_{i=1}^{N-1} \Big (\kappa
|\psi^+_i(x)|^2 \sum_k|\psi^+_k(x)|^2  +\psi_i^{+'}(x) \bar \psi_i^{+'}(x)
\Big )-  2 i \kappa \sum_i \psi_i^+(x_0) \bar \psi_i^{+'}(x_0)
\cr && -2i\kappa \sum_{j} (\psi_j(0)\bar \psi(0))' + {\cal D}_3 + \bar {\cal D}_3  \cr
&& \label{first2}
\ea
where ${\cal D}_{1, 3}$ are defined in (\ref{dd}), we also define: $\bar {\cal D}_3 = \bar D_3 - \bar D_1 \bar D_2 +{\bar D_1^3 \over 3}$ and
\ba
&& \bar D_1 = \kappa {\mathbb P}_{NN} \cr
&& \bar D_2 =  \sum_{j\neq N} W_{Nj}^{-(1)}W_{jN}^{-(1)} + \kappa \sum_{j\neq N} W_{Nj}^{-(1)}{\mathbb P}_{jN} -\kappa \sum_{j\neq N} {\mathbb P}_{Nj} W_{jN}^{+(1)}- \sum_{j\neq N} W_{Nj}^{-(1)}W_{jN}^{+(1)} + \kappa^2 \sum_{j} {\mathbb P}_{Nj} {\mathbb P}_{jN} \cr
&& \bar D_3 =- \sum_{j \neq N} W_{Nj}^{-(1)}W_{jN}^{-(2)}-\sum_{j \neq N} W_{Nj}^{-(2)}W_{jN}^{-(1)} + \sum_{i\neq N  i, j \neq N} W_{Ni}^{-(1)} W_{ij}^{-(1)}W_{jN}^{-(1)}  + \sum_{j \neq N} W_{Nj}^{-(1)}W_{jN}^{+(2)} \cr
&&+ \kappa \sum_{j \neq N} {\mathbb P}_{Nj} W_{jN}^{+(2)}
 -\sum_{i\neq N, j \neq N} W_{Ni}^{-(1)} W_{ij}^{-(1)}W_{jN}^{+(1)} + \sum_{j\neq N} W_{Nj}^{-(2)}W_{jN}^{+(1)} -\sum_{i\neq N,j\neq N} W_{Nj}^{-(1)}{\mathbb P}_{ij} W_{jN}^{+(1)} \label{dd} \cr
&& -\kappa \sum_{j\neq N}W_{Nj}^{-(2)}{\mathbb P}_{jN} + \kappa \sum_{i\neq N,\ j}W_{Ni}^{-(1)} W_{ij}^{-(1)} {\mathbb P}_{jN} + \kappa^2 \sum_{i\neq N, j}W_{Ni}^{-(1)} {\mathbb P}_{ij} {\mathbb P}_{jN} + \kappa^3\sum_{i, j}{\mathbb P}_{Ni} {\mathbb P}_{ij} {\mathbb P}_{jN} \cr && - \kappa^2 \sum_{ j\neq N, i}{\mathbb P}_{Ni} {\mathbb P}_{ij} W^{+(1)}_{jN}.
\ea
We get from the latter the number of particles of the system and the Hamiltonian:
\be
{\cal N} = -{{\cal I}_1 \over 2i \kappa}, ~~~~{\cal H} = - {{\cal I}_3 \over 2i \kappa}.
\ee

It is clear that the presence of non-trivial boundaries yields drastic effects in the bulk behaviour (see the defect contribution in each case, in particular in the SP case note the appearance of the non-trivial $\bar {\cal D}_3$ term!). And although in the SNP case the defect contribution in the momentum is the same as in the periodic case, in the SP case the defect contribution in addition to the ${\cal D}_3$ term includes the $\bar {\cal D}_3$, which is somehow obtained from ${\cal D}_3$ through ``reflection'', compatible with the underlying classical reflection algebra, plus some extra terms associated to the defect degrees of freedom, due to the ${\mathbb L}^{-1}$ contributions.

\section{Discussion}

The study of point like defects in the context of the $\mathfrak{gl}_N$ NLS model via the algebraic description was presented. In addition to the local integrals of motion and the Lax pair, the relevant sewing conditions across the defect point are also derived as analyticity conditions on the time component of the Lax pair. The equations of motion in the bulk and on the defect point are also identified. This is a model associated to the $\mathfrak{gl}_N$ algebra, and the intriguing point, in addition to the defect behavior per se, is that one can implement two distinct sets of boundary conditions and investigate the corresponding modifications in the bulk behavior of the system. Indeed, it turns out that these boundary conditions may drastically alter the bulk behavior, and in particular the defect contribution.

Similar investigations in the context of affine Toda field theories would be also of particular interest. More precisely, identification of the defect contribution in the local integrals of motion as well as the construction of the time component of the Lax pair in the presence of point-like integrable defects would be the main aims of a relevant future study. Again these structures are expected to be modified compared to the periodic case due to the presence of the non-trivial boundaries (see e.g. \cite{avan-doikou-toda, avan-doikou-toda2}).

A similar construction of the time Lax pair component would also provide a significant consistency check in the case of the vector NLS model. It is quite that the presence of the boundary conditions will alter the structure of the time component of the Lax pair in particular in the SP case as is manifest from the form of the associated integrals of motion. The time component defect contribution will be somehow ``doubled'' compared to the periodic case. The two distinct terms contributing to the defect part are expected to be associated via ``reflection'' as is already transparent when computing the Hamiltonian of the system. Hopefully, all the aforementioned issues will be addressed in forthcoming publications.

\end{document}